\documentclass[preprint,showpacs,preprintnumbers,amsmath,amssymb]{revtex4}

\usepackage{graphicx}
\usepackage{dcolumn}
\usepackage{bm}

\begin{document}

\def\be{\begin{equation}}
\def\ee{\end{equation}}
\def\bea{\begin{eqnarray}}
\def\eea{\end{eqnarray}}

\preprint{ } \vskip .5in
\title{Radion effective potential in brane gas cosmology}

\author{Jin Young Kim\footnote{Electronic address:
jykim@kunsan.ac.kr}}
\address{Department of Physics, Kunsan National University,
Kunsan 573-701, Korea}
\date{\today}

\begin{abstract}

We consider a cosmological solution which can explain anisotropic
evolution of spatial dimensions and the stabilization of extra
dimensions in brane gas formalism. We evaluate the effective
potentials, induced by brane gas, bulk flux and supergravity
particles, which govern the sizes of the observed three and the
extra dimensions. It is possible that the wrapped internal volume
can oscillate between two turning points or sit at the minimum of
the potential while the unwrapped three dimensional volume can
expand monotonically. Including the supergravity particles makes
the effective potential steeper as the internal volume shrinks.

\end{abstract}



\maketitle

\section{Introduction}

The idea that the space-time might be more than four dimensions is
being considered widely from particle physics to cosmology.
Obviously string theory provided a strong motivation for
considering higher dimensions. Recently the development of string
theory have led us to diverse cosmological scenarios, for example,
D-brane inflation, moduli inflation, cyclic and ekpyrotic
scenarios, mirage cosmology, and string or brane gas cosmology.
One of the primary goals of string cosmology is to achieve string
compactification which can produce inflation successfully.

Early in the study of cosmology based on string theory, it was
realized that the presence of a gas of strings plays an important
role in the evolution of the universe in Refs. \cite{bv,tv}. They
suggested a mechanism to generate dynamically the spatial
dimensionality of spacetime and to explain the problem of initial
singularity. With the symmetry of string theory, called T-duality,
spacetime has the topology of a nine-dimensional torus, and its
dynamics is driven by a gas of fundamental strings. In string
theory the winding modes can annihilate the anti-winding modes if
they meet. Once the windings are removed from some dimensions,
these dimensions can expand. Since strings have (1+1)-dimensional
world volumes, they can intersect efficiently in 2(1+1) spacetime
dimensions or less. Thus, three spatial dimensions can become
large with a gas of strings.

A gas of fundamental strings was generalized to a gas of various
branes to accommodate the development of D-branes in string theory
\cite{magrio,psl,abe}. Many studies on this issue of string
cosmology with string or brane gas was followed (see
\cite{bW0510022} and references therein for comprehensive
reviews). The key point of replacing string gas with brane gas is
that a hierarchy of scales can be achieved between wrapped and
unwrapped dimensions. It has been checked whether the unwrapped
configuration of branes can generate the inflation successfully
\cite{bem,biswas,campos1,bbst}. Also it is known that string or
brane gases of purely winding modes are not enough to stabilize
the extra dimensions. For example, in eleven dimensional
supergravity, Easter {\it et al.} \cite{egjk} have succeeded in
producing anisotropic expansion by selecting a certain wrapping
matrix. However the radions (scales of the extra dimensions) were
not stabilized.

Stabilization of the radion has been the focus of research in
string or brane gas cosmology
\cite{wb0307044,pb,watson,rador0504047,chatrabhuti,kim0403096,cwb,
kaya,bbc,patil,cw,bw0403075,ks,rador0701029,akk}. One way to
obtain the stabilization of the extra dimensions is to introduce
bulk fields \cite{alexander,fgpt,fgm,campos2}. In the previous
work of the author\cite{kim0608131}, it is shown that the extra
dimensions can be stabilized by including a bulk Ramond-Ramond
(RR) flux in the brane gas formalism. For the specific
configuration of brane gas and RR flux where effectively
six-dimensional brane gas is wrapping the extra dimensions and
4-form RR flux is in the unwrapped dimensions, the flux can cause
a logarithmic bounce to the effective potential as the volume of
the extra dimensions shrinks.

Considering the quantum aspect of the string or brane gas, there
will be a large amount of energy when winding modes and
anti-winding modes of branes annihilate with each other. For
example, string winding modes and anti-winding modes can
annihilate into unwound closed string loops which can be treated
as supergravity particles or radiation. Thus it will be
interesting to see how the simplified stabilization mechanism by
brane gas and flux can be modified if we include supergravity
particles. The purpose of this paper is to extend the previous
study by including supergravity particles.

\section{Brane gas dynamics with flux and supergravity particles}

We consider the ten-dimensional supergravity after the dilaton is
stabilized. We start from the point that winding modes are
annihilated in three spatial dimensions causing them free to
expand while the brane gas remains in the extra six dimensions by
the mechanism of Brandenberger and Vafa \cite{bv}. With a gas of
branes only, the extra dimensions will shrink to zero size and our
assumptions of the brane gas cosmology is not valid anymore. To
prevent this collapse we consider the RR flux in the transverse
dimensions. Thus the bulk effective action consists of graviton
and gauge fields representing the four-form RR flux. If we
consider only the bosonic sector, the effective action can be
written as \cite{kim0608131}
 \be
 S = \frac{1}{2 \kappa^2} \int d^{D+1}x \sqrt{-g} \Bigl( R
 - \frac{1}{2 \cdot 4!} F_4^2 \Bigr), \label{effact}
 \ee
where $R$ is the scalar curvature, $F_4$ is the field strength of
the bulk gauge field, $D=9$, and $\kappa$ is the ten-dimensional
gravitational constant $\kappa^2 = \frac{1}{M^{D-1}_*}$ with $M_*$
being the $(D+1)$-dimensional unification scale.

The gravitational equations of motion are given by
 \be
 G^{MN} = - \kappa^2 ( T_g^{MN} + T_m^{MN} ),  \label{einsteineq}
 \ee
 \be
 \nabla_M F^{MNIJ} = 0,  \label{gaugeeq}
 \ee
where $T_g^{MN}$ is the energy-momentum tensor from four-form
RR-flux
 \be
 T_g^{MN} = \frac{1} {12 \kappa^2} ( F^M_{~~IJK} F^{NIJK}
 - \frac{1}{8} g^{MN} F_{IJKL} F^{IJKL} ) , \label{emtgauge}
 \ee
and $T_m^{MN}$ is the averaged energy-momentum tensor coming from
all the other matter contributions. Also we have the Bianchi
identity since $F$ is an exact form
 \be
 \nabla_{[I} F_{JKLM ]} = 0 . \label{Bianchi}
 \ee

We assume that background fields and matter sources are
homogeneous within the dimensions where they exist. Then we can
treat them as functions of time only. Considering the spatial
section to be a $D$-dimensional torus $T^D$, we write the metric
as
 \be
 ds^2 = - dt^2 + \sum_{k=1}^D a^2_k(t) (dx^k)^2 ,
 ~~~~(0 \le x_k \le 1).
 \label{metric}
 \ee
The non-vanishing components of the Einstein tensor are
 \bea
 G^t{}_t & = & \frac {1}{2} \sum_{k \not= l} \frac{\dot{a}_k}{a_k}
 \frac {\dot{a}_l}{a_l},  \\
 G^i{}_i & = & \sum_{k \not= i} \frac {\ddot{a}_k}{ a_k}
 + \frac{1}{2} \sum_{k \not= l} \frac{\dot{a}_k }{a_k} \frac {\dot{a}_l}{ a_l}
 - \sum_{k \not= i} \frac{\dot{a}_k}{a_k} \frac{\dot{a}_i}{a_i}.
 \eea
As in \cite{kim0608131}, we assume that the RR-flux is confined to
$(3+1)$-dimensional submanifold
 \be
 F^{\alpha\beta\gamma\delta} =
 \frac {\epsilon^{\alpha\beta\gamma\delta} }{\sqrt{-g_4} } F(t),
 \ee
where the Greek indices belongs to $\{ 0, 1, 2, 3 \}$ and $g_4$ is
the induced metric on the $(3+1)$-dimensional submanifold.

With this choice, the Bianchi identity is automatically satisfied
and the solution for $F(t)$ is given by
 \be
 F(t) = \frac{2Q a_1 a_2 a_3 }{ V}, \label{fsol}
 \ee
where $Q$ is an integration constant and $V$ is the total spatial
volume $V = \prod_{k=1}^D a_k$. Then the components of the
energy-momentum tensor by the RR-flux are calculated as
 \bea
 (T_g)_{~t}^t &=& - \frac{1}{\kappa^2} \Bigl( \frac{F(t)}{2}
 \bigr)^2, \\
 (T_g)_{~1}^1 &=& (T_g)_{~2}^2 = (T_g)_{~3}^3
 = - \frac{1}{\kappa^2} \Bigl( \frac{F(t)}{2} \bigr)^2,  \\
 (T_g)_{~4}^4 &=& \cdots = (T_g)_{~D}^D
 =  \frac{1}{\kappa^2} \Bigl( \frac{F(t)}{2} \bigr)^2.
 \eea
This corresponds to the energy momentum tensor of a fluid with
 \be
 \rho_g = \frac{1}{\kappa^2} \Bigl( \frac{F(t)}{2} \bigr)^2,
 ~~~ p_g^1 = p_g^2 =p_g^3 = -\rho_g, ~~~
 p_g^4 = \cdots = p_g^D = \rho_g .  \label{fluxpressure}
 \ee

For other sources of matter, firstly we consider the massless
supergravity particles present in the early universe. The effect
of this source can be expressed by a gas with energy density
$\rho_s$ and pressure $p_s$. We take the gas to be homogeneous and
isotropic perfect fluid with the equation of state
 $p_s = \rho_s /D$. The corresponding energy-momentum tensor is
 \bea
 (T_s)_{~t}^t &=& - \rho_s , \\
 (T_s)_{~1}^1 &=& (T_s)_{~2}^2 =  \cdots = (T_s)_{~D}^D
 = p_s.   \label{sugrapressure}
 \eea
 If we assume that this energy-momentum tensor is covariantly conserved $\nabla_M
T_s^{MN} = 0 $, the energy density scales with time as
 \be
 \rho_s (t) = \rho_s^0 \Big( \frac{V_0}{V(t)}
 \Big)^{\frac{D+1}{D} } ,
 \ee
where $\rho_s^0$ and $V_0$ are the energy density and spatial
volume at time $t_0$.

The second source of matter comes from a gas of branes, wrapped on
the various cycles of the torus. The matter contribution of a
single $p$-brane to the action, if the dilaton is stabilized, is
represented by the Dirac-Born-Infeld (DBI) action
 \be
S_{\rm p} = - T_p \int d^{p+1} \xi \sqrt{ - {\rm det} ( {\hat
g}_{\mu\nu} + {\hat B}_{\mu\nu} + 2 \pi \alpha^\prime {F}_{\mu\nu}
) } , \label{pbract}
 \ee
where $T_p$ is the tension of $p$-brane and ${\hat g}_{\mu\nu}$ is
the induced metric to the brane
 \be
 {\hat g}_{\mu\nu} = g_{MN} \frac{\partial X^M}{ \partial \xi^\mu}
\frac{\partial X^N}{ \partial \xi^\nu}.
 \ee
Here $M,N$ are the indices of $(D+1)$ dimensional bulk spacetime
and $\mu,\nu$ are those of brane. ${\hat B}_{\mu\nu}$ is the
induced antisymmetric tensor field and ${F}_{\mu\nu}$ is the field
strength tensor of gauge fields $A_\mu$ living on the brane. The
fluctuations of the brane coordinates and other fields within the
branes are negligible when the temperature is low enough and the
radii is grown enough. So we neglect ${\hat B}_{\mu\nu}$ and
${F}_{\mu\nu}$ terms. Ignoring the running of the dilaton, we have
absorbed the effect of constant dilaton into the redefinition of
brane tension so that the Einstein frame and the string frame are
equivalent.

We start, after the thermal stage in the early universe by the
mechanism of Brandenberger and Vafa, from the moment that three
dimensions are completely unwrapped. We take the three dimensions
in which the RR flux exists as the unwrapped ones. The other
$(D-3)$ dimensions are wrapped with gases of branes whose
dimensions are less than or equal to $(D-3)$. Assuming each types
of brane gases makes a comparable contribution, we consider a gas
of effective $(D-3)$-branes whose tension we denote by $T_{D-3}$.
Then the energy-momentum tensor for a gas of these branes can be
written as
 \bea
 (T_B)_{~t}^t &=& - \frac{T_{D-3}}{a_1 a_2 a_3} ,  \\
 (T_B)_{~1}^1 &=& (T_B)_{~2}^2 = (T_B)_{~3}^3 = 0,  \\
 (T_B)_{~4}^4 &=& \cdots = (T_B)_{~D}^D
 = - \frac{T_{D-3}}{a_1 a_2 a_3} .
 \eea

Since the $SO(D)$ Poincare invariance is broken down to $SO(3)
\times SO(D-3)$ by RR-flux and $(D-3)$-brane gas, we denote the
scale factor of three dimensional space by $a$ and
$(D-3)$-dimensional subspace by $b$. Then the density and pressure
of the brane gas can be written as
 \be
 \rho_B = \frac{T_{D-3}}{a^3} ,~~~
 p_B^1= p_B^2= p_B^3 = 0, ~~~
 p_B^4 = \cdots =  p_B^D
 = - \frac{T_{D-3}}{a^3} .   \label{branepressure}
 \ee

Finally we include the cosmological constant term which can be
interpreted as the space filling branes
 \be
 (T_{\Lambda})_{~N}^M = {\rm diag} ( -\rho_\Lambda, p_\Lambda ) ,
 \ee
where $\rho_\Lambda = \Lambda$ and $p_\Lambda = - \Lambda$.

Now we insert the energy-momentum tensors, $(T_g)^{MN}$ and
$(T_m)^{MN} = (T_s + T_B + T_{\Lambda})^{MN}$, into the right hand
side of the Einstein equation (\ref{einsteineq}). After some
algebra, the time component of the Einstein equation can be
expressed as, taking units in which $2 \kappa^2 = 1$ for
simplicity,
 \be
 6(\frac{\dot a}{a})^2 + (D-3)(D-4) (\frac{\dot b}{b})^2 + 6(D-3)
 \frac{\dot a}{a} \frac{\dot b}{b} = \rho_s + 2
 \frac{Q^2}{b^{2(D-3)}} + \frac{T_{D-3}}{a^3} + \Lambda .
 \label{eomzerozero}
 \ee
 The spatial components for $SO(3)$ and $SO(D-3)$ subspaces are given by
 \be
 \frac{\ddot a}{a} + 2(\frac{\dot a}{a})^2 + (D-3)
 \frac{\dot a}{a} \frac{ \dot b}{b}
 = -\frac{2(D-4)}{D-1} \frac{Q^2}{b^{2(D-3)}}
 + \frac{\rho_s}{2D}
 + \frac{D-2}{2(D-1)} \frac{T_{D-3}}{a^{3}}
 + \frac{\Lambda}{D-1}, \label{eomso3a}
 \ee
 \be
 \frac{\ddot b}{b} + (D-4)(\frac{\dot b}{b})^2
 + 3 \frac{\dot a}{a} \frac{ \dot b}{b}
 = \frac{6}{D-1} \frac{Q^2}{b^{2(D-3)}}
 + \frac{\rho_s}{2D}
 - \frac{1}{2(D-1)} \frac{T_{D-3}} { a^{3}}
 + \frac{\Lambda}{D-1} . \label{eomso6b}
 \ee

 The key parameters controlling the relative rates of the growth
 for $a$ and $b$ are their accelerations not their velocities. For
 the configuration that $a$ exceeds $b$ by many orders of
 magnitudes, the source terms in Eqs (\ref{eomso3a}) and
 (\ref{eomso6b}) must produce slow-roll conditions for $b$ making
 its acceleration small or negative, while keeping the
 acceleration of $a$ positive.

\section{Effective potential}

 To study the functional behavior analytically, we rewrite the equations of
  motion in terms of $\zeta = a^3$ and $\xi = b^{D-3}$ corresponding to the
 volumes of $SO(3)$ and $SO(D-3)$ subspaces. The constraint equation
 (\ref{eomzerozero}) can be written as
  \be
  \rho_s = \frac{2}{3} (\frac{\dot \zeta}{\zeta})^2
  + \frac{D-4}{D-3} (\frac{\dot \xi}{\xi})^2 + 2
 \frac{\dot \zeta}{\zeta} \frac{\dot \xi}{\xi} - 2
 \frac{Q^2}{\xi^2} - \frac{T_{D-3}}{\zeta} - \Lambda .
 \label{constrainteq}
 \ee
 Using (\ref{constrainteq}), the second derivative equations can be
 written as
  \bea
  \frac{1}{3} \frac{\ddot \zeta}{\zeta} &+& \frac{D-3}{3D} \frac{\dot \zeta}{\zeta}
  \frac{\dot \xi}{\xi} = \frac{1}{3D} (\frac{\dot \zeta}{\zeta})^2
  + \frac{D^2-3D+1}{2D(D-1)} \frac{T_{D-3}}{\zeta}   \nonumber \\
  &+& \frac{D-4}{2D(D-3)} (\frac{\dot \xi}{\xi})^2
  + \frac{-2D^2 + 7D+1}{D(D-1)}\frac{Q^2}{\xi^2}
  + \frac{D+1}{2D(D-1)} \Lambda,  \label{zetatwodoteq}
  \eea
  \bea
  \frac{1}{D-3} \frac{\ddot \xi}{\xi} &+& \frac{3}{D(D-3)} \frac{\dot \zeta}{\zeta}
  \frac{\dot \xi}{\xi} = \frac{1}{3D} (\frac{\dot \zeta}{\zeta})^2
  + \frac{-2D+1}{2D(D-1)} \frac{T_{D-3}}{\zeta}   \nonumber \\
  &+& \frac{D-4}{2D(D-3)} (\frac{\dot \xi}{\xi})^2
  + \frac{5D+1}{D(D-1)}\frac{Q^2}{\xi^2}
  + \frac{D+1}{2D(D-1)} \Lambda.  \label{xitwodoteq}
  \eea
Removing the coupled first derivative terms ($\frac{\dot
\zeta}{\zeta} \frac{\dot \xi}{\xi}$), we have
  \bea
  &&\frac{1}{D-3} \Big\{
  \frac{\ddot \zeta}{\zeta} + \frac{D-6}{9} (\frac{\dot \zeta}{\zeta})^2
  - \frac{9(D^2-3D+1) + (D-3)^2(2D-1)} {6D(D-1)} \frac{T_{D-3}}{\zeta}
  \Big\}   \nonumber   \\
  && = \frac{1}{3} \Big\{ \frac{\ddot \xi}{\xi}
  - \frac{(D-6)(D-4)}{2(D-3)^2} (\frac{\dot \xi}{\xi})^2
  - \frac{(D-3)^2 (5D+1) + 9(2D^2 - 7 D -1)} {D(D-1)(D-3)} \frac{Q^2}{\xi^2}
  \nonumber  \\
  && - \frac{(D-6)(D+1)}{2(D-1)(D-3)} \Lambda  \Big\} .
  \label{separationofvariable}
  \eea

Since the left-hand side is a function of $\zeta$ and the
right-hand side is a function of $\xi$, we take the simplest case
by equating them to a constant $E$ to decouple the variable
$\zeta$ and $\xi$
  \be
  \frac{\ddot \zeta}{\zeta} + \frac{D-6}{D} (\frac{\dot \zeta}{\zeta})^2
  - \frac{9(D^2-3D+1) + (D-3)^2(2D-1)} {6D(D-1)} \frac{T_{D-3}}{\zeta}
  -(D-3)E = 0, \label{eomzeta}
  \ee
  \bea
  \frac{\ddot \xi}{\xi}
  - \frac{(D-6)(D-4)}{2(D-3)^2} (\frac{\dot \xi}{\xi})^2
  &-& \frac{(D-3)^2 (5D+1) + 9(2D^2 - 7 D -1)} {D(D-1)(D-3)} \frac{Q^2}{\xi^2}
  \nonumber  \\
  &-& \frac{(D-6)(D+1)}{2(D-1)(D-3)} \Lambda - 3 E = 0 .
  \label{eomxi}
  \eea
Putting $D=9$, we have
  \be
  \frac{\ddot \zeta}{\zeta} + \frac{1}{3} (\frac{\dot \zeta}{\zeta})^2
  - \frac{41} {16} \frac{T_6}{\zeta}
  - 6E = 0, \label{eomzetad9}
  \ee
  \be
  \frac{\ddot \xi}{\xi}
  - \frac{5}{24} (\frac{\dot \xi}{\xi})^2
  - \frac{47} {8} \frac{Q^2}{\xi^2}
  - \frac{5}{16} \Lambda - 3 E = 0 .
  \label{eomxid9}
  \ee
We remove the first-order derivative terms with $\zeta = f^{3
\over 4}$, $\xi = g^{24 \over 19}$, then the equations reduce to
the motions of a particle with unit mass in one dimension
 \be
 {\ddot f} - \frac{41}{12} T_6 f^{1 \over 4} - 8 E f = 0,
 \label{eomf}
 \ee
 \be
 {\ddot g} - \frac{893}{192} Q^2 g^{- \frac{29}{19}}
 - {19 \over 24} (3 E + {5 \over 16} \Lambda ) g = 0.
 \label{eomg}
 \ee

Now we can analyze the behavior of the two subvolumes by
considering the effective potential as in \cite{kim0608131},
 $ {\ddot f} = - \frac{d V_{\rm eff} (f) }{df}$,
 $ {\ddot g} = - \frac{d V_{\rm eff} (g) }{dg}$.
 The effective potentials are calculated as
 \be
  V_{\rm eff} (f)
 = - \frac{41}{15} T_6 f^{5 \over 4} - 4 E f^2 ,
 \ee
 \be
  V_{\rm eff} (g)
 =  \frac{16967}{1920} Q^2
 g^{- \frac{10}{19}}
 - \frac{19}{48} (3 E + \frac{5}{16} \Lambda ) g^2 .
 \ee
To make the $SO(3)$ subvolume become large indefinitely, $E$ must
be positive and the shape of $V_{\rm eff} (f)$ is given in Fig. 1.
\begin{figure}
\includegraphics[angle=270 , width=8cm ]{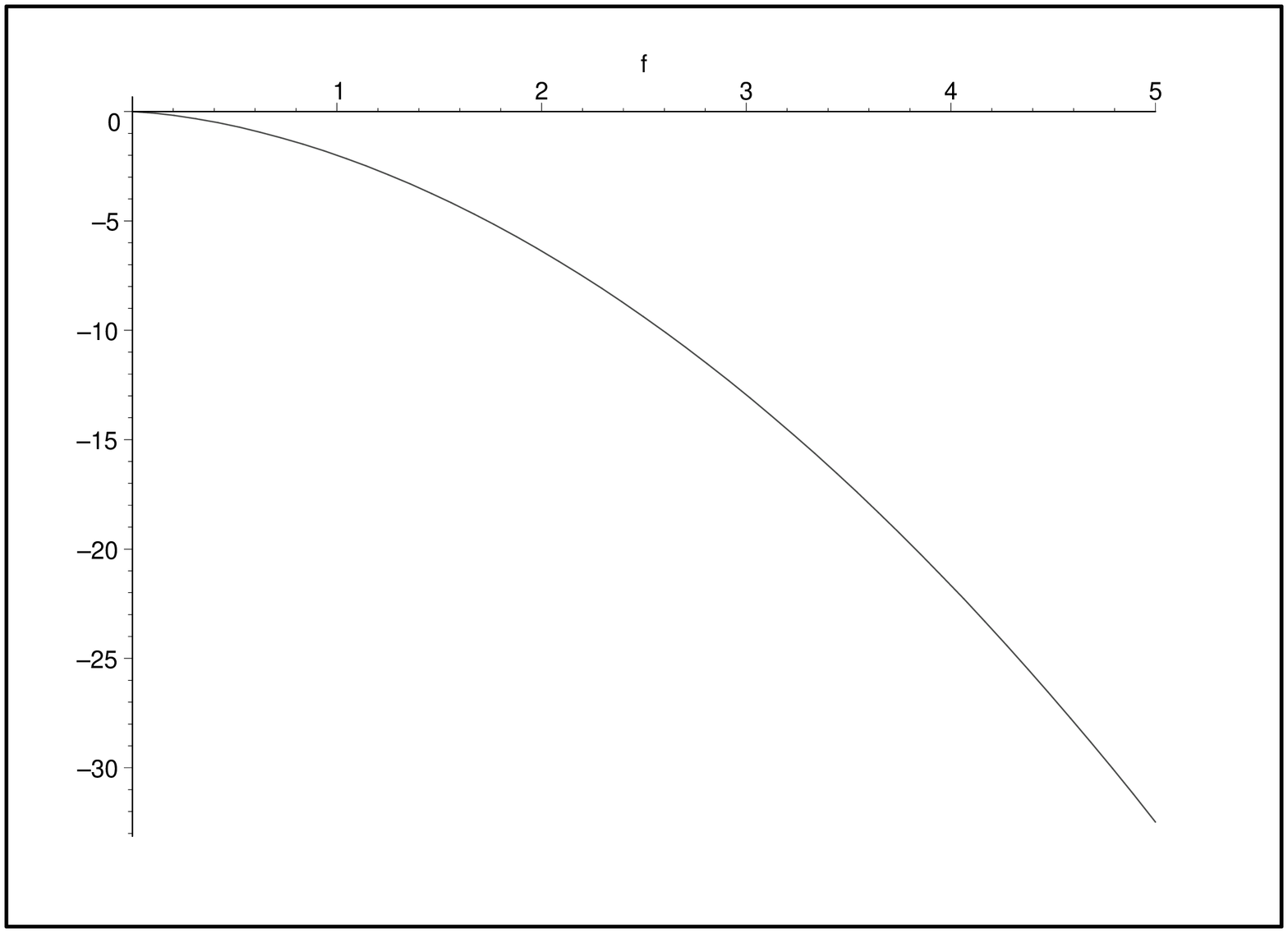} \caption{
Typical shape of effective potential $V_{\rm eff} (f)$ for
unwrapped subvolume for $E > 1$. The plot is for $T_6 = 15/41$ and
$E=1/4$.} \label{fig1}
\end{figure}
The behavior of the effective potential for the $SO(6)$ subvolume
shows a bouncing behavior ($Q^2$ term) for small values of $g$. So
the overall shape of the potential depends on the sign of $3 E +
\frac{5}{16} \Lambda $. For $3 E + \frac{5}{16} \Lambda > 0$, the
shape is given in Fig. 2.
\begin{figure}
\includegraphics[angle=270 , width=8cm ]{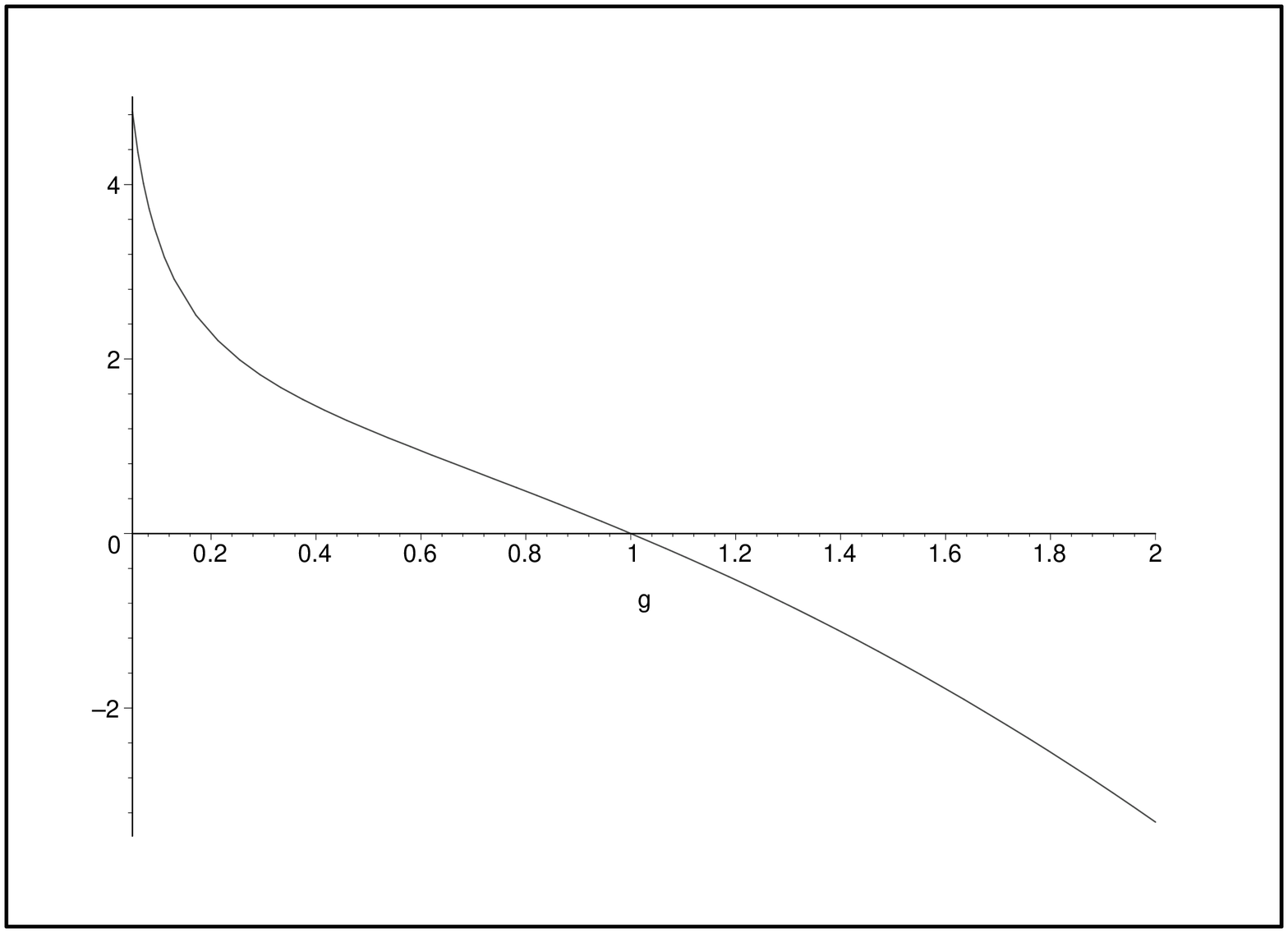} \caption{
Typical shape of effective potential $V_{\rm eff} (g)$ for
 $3 E + \frac{5}{16} \Lambda > 0$. The plot is for $Q^2 = 1920/16967$ and
$3 E + \frac{5}{16} \Lambda = 48/19$.} \label{fig2}
\end{figure}
In this case, both the unwrapped three dimensions and the wrapped
six dimensions expand monotonically. For $3 E + \frac{5}{16}
\Lambda < 0$, the shape is given by Fig. 3.
\begin{figure}
\includegraphics[angle=270 , width=8cm ]{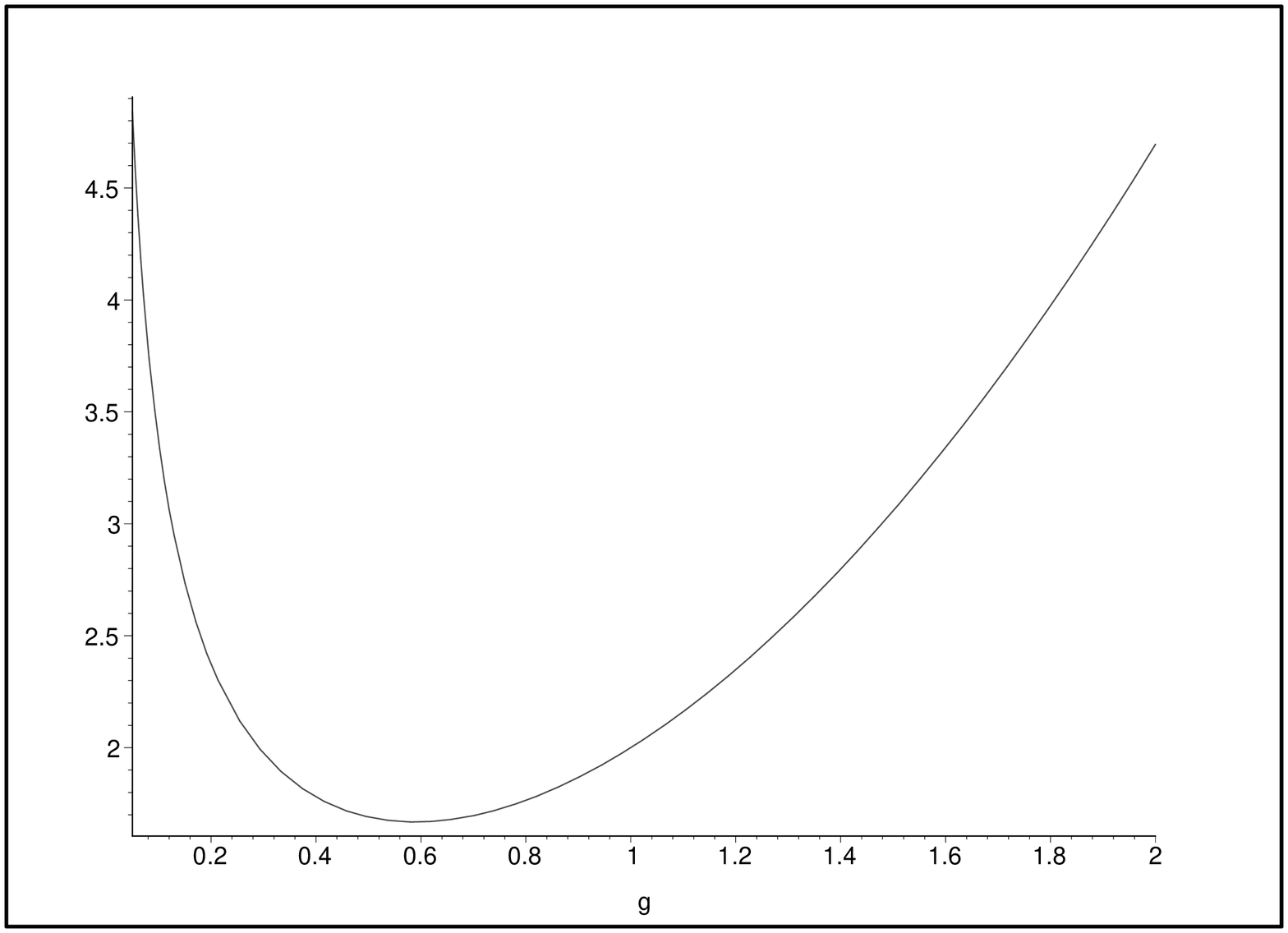} \caption{
Typical shape of effective potential $V_{\rm eff} (g)$ for
 $3 E + \frac{5}{16} \Lambda < 0$. The plot is for $Q^2 = 1920/16967$ and
$3 E + \frac{5}{16} \Lambda = - 48/19$.} \label{fig3}
\end{figure}
In this case, the wrapped internal subvolume can oscillate between
two turning points or sit at the minimum of the potential $g_{\rm
min}$ while the unwrapped subvolume expands indefinitely.

In \cite{kim0608131}, the existence of $(3+1)$-form RR flux in the
unwrapped subspace induces a logarithmic bounce to the effective
potential for small values of $\xi$ and this term prevents the
internal subvolume from collapsing. Including the supergravity
particles into the analysis makes the bounce steeper than the case
with only RR flux. The reason can be understood by looking the
signs of the pressures in Eqs. (\ref{fluxpressure}),
(\ref{sugrapressure}), and (\ref{branepressure}). The brane gas
wrapping the internal dimensions exerts negative pressure and
makes the internal subvolume to contract. However, RR flux and
supergravity particles exert positive pressure to prevent the
internal subvolume from collapsing. The internal volume can be
stabilized by the competition of positive and negative pressures
and our result realized this possibility.

\section{Conclusion}

We have studied the anisotropic evolution of spatial dimensions
and the stability of the extra dimensions with particular emphasis
on the role of supergravity particles. We took a perfect fluid
form for their energy-momentum tensor which drives expansion.
Assuming dilaton is stabilized, we focused on the late time
behavior of brane gas cosmology where windings of branes are
completely removed from three dimensions and RR flux exists in
these unwrapped dimensions. We investigated how the existence of
supergravity particles affects the asymmetric evolution of the
universe by reducing the Einstein equations to the motions of a
particle in one-dimensional effective potential. The shape of the
potential for the three dimensional subvolume is barrier-type so
that it can expand indefinitely. However the shape of the
potential for the extra dimensional subvolume can be well-type so
that it can oscillate between two turning points or be fixed at
the minimum of the potential.

In most approaches to the stabilization in string cosmology, it is
crucial that the dilaton runs to a weak coupling. The scale factor
in the Einstein frame is a linear combination of string frame
dilaton and scale factors. If the dilaton is not fixed, this can
cause serious problems at late-time evolution of the universe
since the Newton constant is not fixed. The extra dimensions can
be unstable as far as the dilaton evolves. It will be an important
challenge to include the running of the dilaton into the
stabilization of the radion.

Recently it is shown that dilaton stabilization and radion
stabilization are compatible by Danos, Frey, and Brandenberger
\cite{dfb}. They identified a stable fixed point corresponding to
the dilaton sitting at the minimum of the potential and the radion
taking on the value at which the enhanced symmetry states are
massless. The stability of this fixed point was analyzed by
studying the linearized equations of motion around the fixed
point. The solution shows a damped oscillatory behavior confirming
the compatibility of two types of moduli stabilization. Despite
the promising result, we have to be very careful when we stabilize
both dilaton and radion simultaneously. Cremonini and Watson
\cite{cw} have discussed the stabilization of moduli in eleven
dimensional supergravity. They found that the production rate of
the BPS bound states could be significant for a modified brane
spectrum with enhanced symmetry. These states can lead to a
stabilization by an attractor mechanism. However, the
stabilization drives the evolution to a region where a
perturbative description of the string dynamics fails. That is,
the supergravity approximation is not valid in this region. It is
important not to forget the string theory origin of the low-energy
effective action.

Realizing the transitions between the different thermodynamic
phases of string gas is important in string cosmology. In
\cite{kalwat}, it is pointed out that the dilaton field may
obstruct the transition from the Hagedorn phase of hot and dense
string gas to expanding FRW phase of dilute string gas. They
categorized the possible branches of the solutions according the
sign of the dimensionally reduced effective dilaton and noticed
that the branch changing is impossible as long as the energy
density of the universe is not negative. Finding the appropriate
energy sources which enable the phase changing seems another
challenge in string/brane gas cosmology.

\begin{acknowledgments}
This work was supported by research funds of Kunsan National
University.
\end{acknowledgments}

\end{document}